# Integrated-Photonics-Based Systems for Polarization-Gradient Cooling of Trapped Ions

Sabrina M. Corsetti[1], Ashton Hattori[1], Ethan R. Clements[1], Felix W. Knollmann[1], Milica Notaros[1], Reuel Swint[2], Tal Sneh[1], Patrick T. Callahan[2], Gavin N. West[1], Dave Kharas[2], Thomas Mahony[2], Colin D. Bruzewicz[2], Cheryl Sorace-Agaskar[2], Robert McConnell[2], Isaac L. Chuang[1], John Chiaverini[1,2], & Jelena Notaros[1,*]

[1] Research Laboratory of Electronics, Massachusetts Institute of Technology, Cambridge, MA 02139, USA

[2] Lincoln Laboratory, Massachusetts Institute of Technology, Lexington, MA 02421, USA

[*] Corresponding Author: notaros@mit.edu

## Abstract

Trapped ions are a promising modality for quantum systems, with demonstrated utility as the basis for quantum processors and optical clocks. However, traditional trapped-ion systems are implemented using complex free-space optical configurations, whose large size and susceptibility to vibrations and drift inhibit scaling to large numbers of qubits. In recent years, integrated-photonics-based systems have been demonstrated as an avenue to address the challenge of scaling trapped-ion systems while maintaining high fidelities. While these previous demonstrations have implemented both Doppler and resolved-sideband cooling of trapped ions, these cooling techniques are fundamentally limited in efficiency. In contrast, polarization-gradient cooling can enable faster and more power-efficient cooling and, therefore, improved computational efficiencies in trapped-ion systems. While free-space implementations of polarization-gradient cooling have demonstrated advantages over other cooling mechanisms, polarization-gradient cooling has never previously been implemented using integrated photonics. In this paper, we design and experimentally demonstrate key polarization-diverse integrated-photonics devices and utilize them to implement a variety of integrated-photonics-based polarization-gradient-cooling systems, culminating in the first experimental demonstration of polarization-gradient cooling of a trapped ion by an integrated-photonics-based system. By demonstrating polarization-gradient cooling using an integrated-photonics-based system and, in general, opening up the field of polarization-diverse integrated-photonics-based devices and systems for trapped ions, this work facilitates new capabilities for integrated-photonics-based trapped-ion platforms.

## Introduction

The intrinsic reproducibility, long coherence times, and strong interactions available in collections of trapped atomic ions have enabled a variety of quantum logic operations with fidelities sufficient for fault-tolerant quantum computing, including single- and two-qubit gates, qubit state preparation, and readout [1–4]. At the system level, trapped ions have been successfully implemented as the basis for quantum processors and high-accuracy optical clocks [5–10]. However, scaling trapped-ion systems to a large number of qubits while maintaining high fidelities remains a significant challenge for achieving practical, general, and portable quantum systems [4,11,12].

Traditionally, trapped-ion systems have been implemented using complex free-space optical configurations, whose large size and susceptibility to vibrations and drift can limit the fidelity and addressability of ion arrays, inhibiting scaling to large numbers of qubits. Recently, integrated-photonics-based devices and systems have been demonstrated as an avenue to address these challenges by enabling photonic routing and optical-beam emission from systems with a scalable chip-based form factor and inherent stability [11–14]. For example, foundational works have developed and utilized low-loss integrated-photonics platforms [15] to enable trapped-ion operations ranging from photoionization and cooling [11] to two-qubit operations [12]. More recent demonstrations have further advanced the field, demonstrating steps towards scaled multi-zone operation of integrated-photonics-based ion-trap chips [13,14].

While these previous demonstrations have utilized integrated-photonics-based beam delivery to implement both Doppler and resolved-sideband cooling, these cooling techniques are respectively limited by a higher final kinetic energy (the Doppler cooling limit) [16,17] and the inherent slowness of addressing individual motional sidebands [18,19]. Overcoming these limitations is of particular interest for improving trapped-ion computational efficiencies, as inefficient ground-state cooling can comprise a significant portion of a trapped-ion system's computational time [20–22]. Compared to Doppler and resolved-sideband cooling, alternative cooling mechanisms provide useful tradeoffs in rate and final temperature to enable faster and more power-efficient cooling and, therefore, improved computational efficiencies for trapped-ion systems. One such higher-efficiency cooling mechanism is polarization-gradient cooling, which offers the potential for parallel cooling of multiple motional modes to below the Doppler limit [23].

Prior demonstrations of polarization-gradient cooling have successfully cooled chains of up to 51 ions to sub-Doppler temperatures over multiple motional modes simultaneously, demonstrating the advantage offered by polarization-gradient cooling [24–27]. However, all existing demonstrations of polarization-gradient cooling to date have been implemented using free-space beam delivery. Accordingly, polarization-gradient cooling has never previously been implemented using integrated photonics. This is primarily due to the stringent beam-delivery requirements imposed by polarization-gradient cooling. While both Doppler and resolved-sideband cooling can be performed using only a single linearly-polarized beam in some atomic species, polarization-gradient cooling relies on the interaction between an atom and a polarization gradient generated by the interference between either two circularly-polarized beams or two beams with orthogonal linear polarizations [23]. To enable such configurations using integrated photonics, multiple emitters must be arranged to simultaneously deliver beams to a target ion location. Furthermore, some configurations require the emission of beams with different polarizations. These configurations necessitate the co-optimization of multiple polarization-diverse emitters for equal-intensity beam formation. Advances have been made in recent years targeting polarization-diverse integrated-photonics devices for qubit interactions [28–30] and multi-beam trapped-ion functionalities have been demonstrated using integrated-photonics-based trap chips, such as simultaneous Doppler cooling and repumping [11]; however, all previous demonstrations of cooling performed by integrated-photonics-based trapped-ion systems have been limited to emission of single transverse-electric-polarized (TE-polarized) beams [11–14], thus inhibiting the implementation of polarization-gradient cooling.

In this paper, we design and experimentally demonstrate key polarization-diverse integrated-photonics devices and utilize them to implement a variety of integrated-photonics-based polarization-gradient-cooling systems, culminating in the first integrated-photonics-based demonstration of polarization-gradient cooling of a trapped ion (characterized in detail in a companion paper [31]). First, we discuss the mechanisms underlying polarization-gradient cooling and introduce several integrated-photonics-based architectures capable of performing polarization-gradient cooling. Second, we design and experimentally demonstrate the key integrated-photonics routing and grating-emitter devices required to implement these architectures. Third, we incorporate these devices into, and successfully experimentally demonstrate, multiple integrated-photonics-based polarization-gradient-cooling systems. Finally, we implement one of these systems within an integrated-photonics-based ion-trap chip and use it to demonstrate the first polarization-gradient cooling of a trapped ion using an integrated-photonics-based system (characterized in detail in [31] and summarized in this work). By demonstrating polarization-gradient cooling using an integrated-photonics-based system and, in general, opening up the field of polarization-diverse integrated-photonics-based devices and systems for trapped ions, this work facilitates new capabilities for integrated-photonics-based trapped-ion platforms.

## Results

**Conceptual overview of integrated-photonics-based architectures for polarization-gradient cooling**

Integrated-photonics-based ion-trap chips build upon traditional surface-electrode-based trap chips by integrating both photonics layers and surface electrodes into a single planar electronic-photonic trap chip [11–14]. The surface electrodes allow for confinement and routing of ions within radio-frequency traps formed over the trap-chip surface [32]. Integrated-photonics layers underneath the electrodes allow for routing of light towards integrated grating couplers that emit optical beams towards the trapped ions through gaps in the surface electrodes, as conceptually depicted in Figs. 1a and 2a.

In prior work [11], individual integrated gratings emitting TE-polarized light were used analogously to linearly-polarized free-space beams to demonstrate Doppler and resolved-sideband cooling of a trapped

$^{88}$Sr$^+$ ion using an integrated-photonics-based system. First, integrated-photonics-based Doppler cooling along the $^{88}$Sr$^+$ S$_{1/2}$ → P$_{1/2}$ transition was enabled using a 422-nm-wavelength beam. The further implementation of integrated-photonics-based resolved-sideband cooling by driving the S$_{1/2}$ → D$_{5/2}$ transition using a 674-nm beam, coupled with pumping along the D$_{5/2}$ → P$_{3/2}$ transition using a 1033-nm beam, enabled cooling of the average axial-mode motional-state occupation to $\bar{n} < 1$. In this work, we extend beyond these previously demonstrated cooling techniques by enabling polarization-gradient cooling using polarization-diverse integrated-photonics-based devices and systems, paving a path towards enhanced trapped-ion state-preparation and computational efficiencies [20].

As introduced above, polarization-gradient cooling is enabled by the interaction of multiple beams with either orthogonal linear polarizations (the lin ⊥ lin configuration) or opposite-handed circular polarizations (the $\sigma^+$-$\sigma^-$ configuration) [23]. In this work, we focus on the lin ⊥ lin configuration. In the ideal case, this configuration is implemented using two counterpropagating beams with wavelength $\lambda$. The orthogonal polarizations of the beams create a polarization gradient of varying ellipticity (between circular and linear polarization) along the propagation axis of the beams. To summarize the theory underlying the cooling mechanism performed by this ideal gradient (as first introduced in [22]), we consider an ion with ground-state Zeeman sublevels $g_{\pm 1/2}$ subject to light shifts induced by the polarization gradient. In regions of pure $\sigma^-$ circular polarization, the $g_{1/2}$ ground state is strongly coupled to the light field and experiences a large light shift, while the $g_{-1/2}$ state remains unaffected. The converse relationship holds for $\sigma^+$ polarization. As a result, the light-shifted energy levels of $g_{\pm 1/2}$ obtain a periodic spatial dependence. By choosing the laser detuning such that the ion is optically pumped to the lower-energy Zeeman sublevel during its motion, we enact motional-state cooling by means of periodic stimulated loss of kinetic energy [23].

In practice, many beam and magnetic-field orientations are conducive to lin ⊥ lin polarization-gradient cooling. Ideally, the polarization vectors and applied magnetic fields comprising any given cooling configuration will be mutually orthogonal, resulting in the formation of a polarization gradient with alternating $\sigma^+$-$\sigma^-$ polarization at the intersection of the beams. Figures 1b-d depict three example polarization-gradient-cooling configurations explored in this work that are compatible with being emitted by a planar trap chip. Since it is challenging to emit counterpropagating beams from a planar trap chip that intersect at an ion location, these configurations consist of orthogonally-polarized beams emitted by the trap chip with a near-45° angle from the vertical such that the beams intersect at a target ion location over the trap-chip surface, as conceptually depicted in Fig. 1a. For each of the configurations in this work, we aim to enable simultaneous cooling of all three modes of a trapped ion's motion by offsetting the polarization gradient from the trap axis by 45° in the plane of the trap chip and applying appropriate static voltages to rotate the combined radio-frequency and static potential such that the polarization gradient has a component along all three principal trap axes.

To physically implement each of these three beam configurations, we integrate two grating emitters into an electronic-photonic trap site, positioned equidistant from a target ion location. Each integrated grating emits a focused 422-nm-wavelength beam at a near-45° angle from the vertical towards the ion located 50 µm over the surface of the chip. These architectures, depicted in Figs. 2b-d, include two TE-polarization-emitting gratings placed orthogonal to each other (TE-TE), two transverse-magnetic-polarization-emitting (TM-polarization-emitting) gratings oriented opposite each other (TM-TM), and a TE- and a TM-polarization-emitting grating placed opposite each other (TE-TM).

**Design of polarization-diverse grating-emitter pairs**

As a central component of these polarization-gradient-cooling architectures, we must design integrated grating emitters capable of emitting intensity-matched beams of diverse polarizations from a planar trap chip. In this section, we design a pair of integrated TE- and TM-emitting gratings with an operating wavelength of 422 nm, corresponding to the $^{88}$Sr$^+$ S$_{1/2}$ → P$_{1/2}$ transition.

To design this polarization-diverse grating pair, we implement a custom design suite to realize bilayer, apodized, and curved gratings that emit unidirectional focused beams that maximize the percentage of light emitted towards the target ion location 50 μm above the surface of the chip. As depicted in Fig. 2a, the gratings are composed of two 100-nm-thick layers of silicon nitride (SiN) separated by 90 nm of silicon dioxide (SiO$_2$). The periods and lengths of each layer's grating teeth are optimized such that the angle of emission varies along the length of the grating to enable focusing in the propagation dimension ($x$), as depicted in Fig. 3a. In addition, the teeth in each layer are offset from each other in the propagation dimension ($x$) to enable unidirectional emission upwards [33], as depicted in Fig. 3b. Finally, the grating teeth are curved to enable focusing in the transverse dimension ($y$) of the grating, as depicted in Fig. 3a. The gratings are fed by a 10-µm-long large-angle free-propagation taper such that the input fundamental waveguide mode freely expands up to the full width of the grating.

To initialize the optimization suite, we begin by defining the physical features and constraints of our gratings: the fabrication stack, minimum fabricable feature size, grating aperture size, target ion location, operating wavelength, and input polarization (TE or TM). Then, using these values as guidelines, the suite determines the optimal grating parameters (upper-layer and lower-layer tooth lengths and layer offset in the $x$ dimension, as labeled in Fig. 3b) for each period within the grating. To perform this optimization, we first define the range of emission angles necessary to enable focusing in the propagation dimension. Then, we perform a parameter optimization for each angle within this range. To begin each optimization, we initialize a particle-swarm algorithm in MATLAB [34] with an initial set of three parameters: upper-layer tooth length, lower-layer tooth length, and layer offset in the $x$ dimension. The algorithm then calls a Lumerical-FDTD electromagnetic-simulation script [35] that constructs a grating with the given geometric parameters. We then simulate the performance of this grating with a given input polarization (TE or TM). Once the simulation is complete, we integrate the field above the grating to determine the fraction of light, $L_{ion}$, emitted at the desired target angle. Next, we compute 1 - $L_{ion}$ - $L_{rem}$, where $L_{rem}$ is the fraction of light remaining at the end of the simulated grating due to incomplete scattering, and feed the value of $L_{ion}$/(1 - $L_{ion}$ - $L_{rem}$) back into the MATLAB algorithm as a figure of merit representing the ratio of desired-angle to undesired-angle scattering. The particle-swarm algorithm then determines a new set of test-parameter values, and the process repeats until an optimal combination is reached. We complete this parameter optimization for all emission angles within our defined angle range (with resulting performance shown in Figs. 3c-d) to generate a dataset that maps the range of desired emission angles to grating parameters. Finally, we utilize this dataset to construct the grating in a tooth-wise fashion, determining the desired emission angle for each grating period given the period's $x$ position and interpolating the dataset to determine the optimal upper-layer and lower-layer tooth lengths and $x$-dimension layer offset for each period.

Following this parameter-selection process, we simulate free expansion of the input waveguide mode within a slab to determine the phase front of the modes propagating into the grating. This allows us to determine the optimal curvature for each grating tooth to allow for lateral-dimension focusing [36]. Finally, we run a full-scale 3D simulation of the designed grating and verify that it performs as intended.

We perform this grating-design process for gratings of multiple different lengths in order to converge on a pair of optimized TE and TM gratings with matching beam intensities at the target ion location, a necessary condition for many ion operations, including the generation of high-purity circular polarization within a TE-TM polarization gradient [23]. Optimizing the grating length is necessary to perform intensity matching between TE and TM gratings due to the fundamentally different scattering behaviors of TE- and TM-polarized modes; as TE waveguide modes are more highly confined than TM modes and, therefore, have higher effective indices of refraction [37], TE-polarized light scatters more strongly than TM-polarized light from grating emitters with a given minimum feature size. For this reason, the maximum effective aperture size of a fabricable TE grating is intrinsically shorter than that of a TM grating, resulting in fundamentally-improved beam-focusing abilities for TM gratings. Thus, an optimal grating length will truncate the effective aperture of the TM grating such that it focuses with an intensity comparable to the TE grating at the ion location. To determine this optimal grating-aperture size, we design multiple TE-TM grating pairs with different lengths and compare the intensities of the gratings in each pair. After completing this process, we choose a final aperture size that results in matched intensities, with a final result of 17 × 18 µm in this particular case. The simulated efficiencies of the resulting intensity-matched TE and TM gratings, defined as the percentage of light concentrated in a small (2 × 2 μm) region surrounding the target ion location, are 18.4% and 18.8%, respectively.

To compensate for potential fabrication-bias-induced deviations in grating performance from simulation, we then design a suite of grating variants. First, we pick a nominal target ion location around which we center our variants. We choose an ion location of $x = 70.7$ μm from the start of the 10-μm-long grating-input tapers at a height of $z = 50$ μm, with our ion height fixed by our radio-frequency trap geometry [31]. For both the TE and TM polarizations, we then design seven 17-μm-long gratings each targeting a slightly different $x$-axis position (61.7 to 79.7 μm, in steps of 3 μm), with all variants assuming a constant $z = 50$ μm ion height. In addition, to compensate for potential fabrication-bias-induced deviations in scattering strength and therefore effective-aperture sizes, we design grating variants with multiple lengths. Specifically, for both the TE and TM polarizations, we design three 15-μm-long gratings (targeting positions of $x = 62.7$ μm, 68.7 μm, and 74.7 μm) and three 19-μm-long gratings (targeting positions of $x = 66.7$ μm, 72.7 μm, and 78.7 μm), with all variants assuming a constant $z = 50$ μm ion height.

The resulting simulated emission profiles in the $xz$ plane for an example pair of characteristic gratings are plotted in Figs. 4a-b. The example TE grating (17 μm long and targeting $x = 79.7$ μm) has an expected emission angle of 50.3° and a spot size ($1/e^2$ diameter) of 9.9 μm × 5.0 μm at the ion location 50 μm above the chip surface. The example TM grating (17 μm long and targeting $x = 76.7$ μm) has an expected emission angle of 46.9° and a spot size of 6.1 μm × 3.7 μm at the ion location 50 μm above the chip surface.

**Experimental demonstration of polarization-diverse grating-emitter pairs**

After completing the grating-design process, we fabricate and optically test our suite of 13 TE and 13 TM gratings to verify their compatibility with our polarization-gradient-cooling architectures.

We fabricate our gratings in a 200-mm wafer-scale fabrication process developed at MIT Lincoln Laboratory for low-loss routing at wavelengths spanning the spectrum from ultraviolet to near infrared [15]. The platform (Fig. 2a) contains three 100-nm-thick waveguiding layers: one in amorphous aluminum oxide ($Al_2O_3$) and two in silicon nitride (SiN), each separated by 90 nm of silicon dioxide ($SiO_2$). The waveguiding layers are clad in 5 μm of $SiO_2$, followed by a niobium (Nb) surface-electrode layer.

To characterize the fabricated gratings (Fig. 5b), we route light from a benchtop laser to the photonics test chip through a series of polarization-maintaining fibers. To set the polarization at the input of the chip, we fix a polarizing beamsplitter cube to a mount aligned with the chip axes. Then, we use transmission through the beamsplitter ports to align the incident polarization of the fiber to the operating polarization of the grating under test (either TE or TM). We then remove the cube and couple light onto the chip using an on-chip $Al_2O_3$ inverse-taper fiber-to-chip edge coupler (Fig. 5c). The light in the $Al_2O_3$ waveguide is then routed through a series of integrated layer-transition escalators: the first transitioning from $Al_2O_3$ to the first SiN waveguiding layer (Fig. 5c) and the second from SiN to dual-layer SiN (Fig. 5c) [15]. Finally, the light is routed to one of the dual-layer grating emitters.

We image the grating emission using a 50X objective and a visible-light camera. To measure each grating's performance, we use an automated setup (Fig. 5a) to increment the height of the imaging optical train in 1-μm steps and capture images of the grating's emitted beam over a range of 0 to 100 μm above the surface of the chip (Figs. 6a-b). We then use the resulting data to compute the grating's emission angle and beam dimensions (Figs. 6c-d). Using this characterization procedure, we find that a characteristic fabricated TE grating (17 μm long and targeting $x = 79.7$ μm) successfully emits TE-polarized light at an angle of approximately 43.2° (Fig. 6a) with a spot size ($1/e^2$ diameter) of 7.6 μm × 4.3 μm at 50 μm above the chip surface (Fig. 6c). Similarly, a characteristic fabricated TM grating (17 μm long and targeting $x = 76.7$ μm) successfully emits TM-polarized light at an angle of approximately 43.1° (Fig. 6b) with a spot size of 5.0 μm × 3.6 μm at 50 μm above the chip surface (Fig. 6d).

As the optimal ion height ($z = 50$ μm) is a fixed property of the chips' radio-frequency trapping potential, it is necessary to determine the offset of the emitted beams from a target ion $x$ position at this height. Using the characterization procedure introduced above, we image the beams emitted by each of the 26 grating variants within the intended ion-height plane at $z = 50$ μm. Within this plane, we determine the $x$ position of the emitted beam and calculate the offset from the target design value. We collect this $x$-position data for all 26 grating variants from five chips sampled from disparate locations across two fabricated wafers (total of 130 gratings measured), to provide a robust analysis of grating performance across sites subject to differing fabrication conditions. The results, plotted in Figs. 6e-f, demonstrate that grating performance remains highly consistent across different chips and wafers. Although consistent, we observe a fabrication-induced bias in our measured $x$ positions. While our nominal TE and TM gratings were both originally designed to target $x = 70.7$ μm, the fabricated nominal TE and TM gratings have measured $x$ positions of approximately $63.2 \pm 0.6$ μm and $64.9 \pm 1.1$ μm, respectively. We expect that this discrepancy is due to fabrication bias reducing the effective index of the modes propagating in the gratings. To compensate for these consistently-measured deviations, we can use alternate grating variants for future ion-trap grating configurations. For example, to target a position of $x = 70.7$ μm, we can select the TE grating originally designed to target $x = 79.7$ μm and the TM grating originally designed to target $x = 76.7$ μm.

Finally, we experimentally characterize the efficiencies of the fabricated gratings. To perform this characterization, we first determine the conversion from raw image-brightness values to optical power for the grating-imaging camera. To do this, we couple light from a 422-nm-wavelength laser diode into an optical fiber and align the fiber's output facet with our visible-light camera. Then, we capture images of the fiber facet at 25 different output powers, as recorded by a power meter, sum the reported pixel-brightness values of the facet images, and fit a line mapping the integrated image brightness values to optical power. Next, to isolate the grating efficiencies from other on-chip loss mechanisms, we must account for losses in the waveguide routing. By repeatedly experimentally measuring fiber-to-chip coupling losses, $Al_2O_3$-to-SiN escalator losses, and SiN-to-dual-SiN escalator losses across multiple chips and wafers, we find an expected total average routing loss of approximately 9.7 dB for TE and approximately 9.3 dB for TM polarization, which we account for in our reported grating efficiencies. Finally, we experimentally measure the power emitted from all 26 grating variants from five chips (total of 130 gratings measured) and calculate their final grating efficiencies, defined as the percentage of light concentrated in a small (2 × 2 μm) region surrounding the beam centroid at $z = 50$ μm. We find that the efficiencies span approximately 5-15% across the different variants, closely matching our simulated efficiency values.

**Experimental demonstration of integrated-photonics-based systems for polarization-gradient cooling**

After characterizing our individual TE and TM gratings, we utilize them as building blocks to experimentally demonstrate our three proposed polarization-gradient-cooling systems.

We implement two different system categories – single- and dual-fiber-input systems – to enable multiple modalities of polarization-gradient cooling. The dual-fiber-input systems with an off-chip splitter facilitate traveling-wave cooling implementations similar to previous free-space demonstrations [24,26,27], whereas the single-fiber-input systems with an on-chip splitter are capable of producing a phase-stable polarization gradient and, consequently, enable cooling at specific phases within this stable gradient for the first time, as demonstrated in the following section.

We begin by introducing and characterizing the single-fiber-input (TE-TE and TM-TM) systems. These systems consist of an on-chip inverse-taper $Al_2O_3$ edge coupler routed to a polarization-specific (TE- or TM-optimized) on-chip $Al_2O_3$ multimode-interferometer (MMI) splitter that splits light from a single input $Al_2O_3$ waveguide into two output $Al_2O_3$ waveguides (Fig. 7a) [38]. These separate waveguides are then routed through the integrated $Al_2O_3$-to-SiN and SiN-to-dual-SiN escalators discussed in the previous section. Finally, each dual-layer arm routes to a single integrated grating emitter, with the grating orientation determined by the desired beam configuration. The gratings are oriented orthogonally in the TE-TE case and opposite each other in the TM-TM case, as shown in Figs. 2b-c. We fabricate these systems using the same 200-mm wafer-scale process introduced in the previous section and depicted in Fig. 2a [15].

To test these systems, we utilize the same setup used in the previous section, shown in Fig. 5a. We once again image the chip with a 50X objective and a visible-light camera, set our input polarization using a free-space polarizing beamsplitter, and use our automated setup to increment the height of the imaging optical train in 1-μm steps, this time capturing the light emitted by both gratings simultaneously. Characteristic side-view and top-down scan results for the TE-TE and TM-TM systems are provided in Figs. 7c-d.

To characterize the effects of the fabrication bias on these systems, we measure the performance of three TE-TE- and three TM-TM-system variants. Each system variant contains a distinct pair of TE- or TM-grating variants. Each grating-variant pair targets a different $x$ position, corresponding to a different angle of emission and, accordingly, a different height of intersection, $z$, for the two-grating systems. Specifically, we target $x = 64.7$ μm, 70.7 μm, and 76.7 μm. By measuring the performance of each of these variants, we can characterize the fabrication-bias-induced deviation of our measured intersection heights from the target design values. We measure these system variants across 8 chips from 3 wafers (total of 48 systems measured) and fit trendlines to the data (Fig. 7b). We find that our linear fits strongly agree with the data presented in the previous section for our single-grating test-structure $x$ positions; to target a beam-intersection height of $z = 50$ μm (corresponding to $x = 70.7$ μm), we can use the TE grating originally designed to target $x = 79.7$ μm and the TM grating originally designed to target $x = 76.7$ μm.

From the top-down views of the single-fiber-input systems, we find that the gratings emit with closely-matched intensities. This is due to the even splitting ratios of the integrated MMI devices; for the TE-optimized and TM-optimized MMIs, we measure splitting ratios of -3.02±0.36 dB and -2.93±0.24 dB, respectively, with the target -3.01 dB ratio within a single standard deviation for both MMIs.

Notably, at the intersection height of each beam pair, we observe an optical-intensity gradient along the polarization-gradient axis. While theory would predict that beams with perfectly orthogonal polarizations will not generate an observable interference pattern, we find that these patterns are due to both the alignment of the magnetic fields of our beams as well as the quasi-TE / quasi-TM nature of waveguide modes [39]. As physical waveguide modes are bounded in three dimensions, they do not exhibit pure TE or TM polarization; instead, waveguide modes exhibit strong TE- or TM-like polarization, with non-zero electric-field contributions along all three principal axes [39]. As a result, beams emitted by waveguide-based devices also generally exhibit multiple non-zero field components. In simulation, we find that the non-orthogonal components of the emitted beams correspond to on the order of 10% of the total optical power in the quasi-TE beams and on the order of 1% of the total optical power in the quasi-TM beams. However, this power contribution alone does not dictate the interference pattern of these systems. For example, for the particular case of the TM-TM systems, the beams are aligned such that their magnetic fields exhibit a near-total overlap. In contrast, the TE-TE systems are aligned such that the dual-beam magnetic-field overlap is significantly reduced. This results in a larger magnetic-field interference contribution for the TM-TM systems than for the TE-TE systems. The final interference patterns generated by each dual-grating system are thus a result of several factors including the angle between the beams, the polarization purity of the emitted waveguide modes, and the angle of emission. Despite the hybrid-polarization emission and intensity interference patterns, we successfully observe cooling, as demonstrated in the following section. We expect that this polarization impurity decreases the purity of our polarization gradient's circular-polarization regions, leading to a decreased difference in the $g_{\pm 1/2}$ light shift and, therefore, reduced cooling efficiency. Further parametrizing our dual-grating-system interference patterns and quantifying the effects of interference on our polarization gradient could be explored further in future work.

After thoroughly characterizing the single-fiber-input polarization-gradient-cooling systems, we expand upon our testing setup to also characterize dual-fiber-input systems. Our dual-fiber-input structures begin with one on-chip inverse-taper $Al_2O_3$ edge coupler on either side of the photonics test chip. Then, each separate $Al_2O_3$ waveguide is routed through the same integrated $Al_2O_3$-to-SiN and SiN-to-dual-SiN escalators discussed above. Finally, each arm terminates in a single grating: one TE- and the other TM-emitting, as shown in Fig. 2d. To test these systems, we route light from a benchtop laser to a single-mode fiber routed through polarization paddles. We then couple this single-mode fiber to a fiber-based polarizing beamsplitter with a single-mode-fiber input and two polarization-maintaining-fiber outputs. Finally, we align one of the two polarization-maintaining-fiber arms to each side of the photonics test chip, set the polarization at the respective inputs to TE and TM using a polarizing beamsplitter cube, and couple light onto the chip using the two on-chip $Al_2O_3$ inverse-taper edge couplers.

Characteristic measurement results for the dual-fiber-input systems are provided in Fig. 7e. While we demonstrate that these dual-fiber-input systems remain sufficiently stable for us to image an interference pattern within the exposure time of the camera, these dual-fiber-input systems are inherently less stable than the single-fiber-input systems demonstrated above, due to increased susceptibility to relative phase variability between the two paths induced by using an off-chip fiber-based, rather than an on-chip, splitter. As a result, the dual-fiber-input systems are better suited for polarization-gradient cooling implementations with a traveling-wave gradient, more similar to previous free-space demonstrations [24,26,27]. To instead achieve the long-term stability offered by single-fiber-input systems, we could implement our TE-TM architecture with an integrated-photonics polarization rotator operating at 422 nm, as we developed and experimentally demonstrated in [40]. Such an architecture would allow us to couple TE-polarized light onto the chip using a single input edge coupler, route this light to an on-chip TE-optimized MMI, route one of the MMI-output arms through a polarization rotator to convert the arm's polarization from TE to TM, then route each arm to a polarization-specific grating, as we outlined in [41].

**Experimental demonstration of polarization-gradient cooling of a trapped ion**

Using the single-fiber-input TE-TE polarization-gradient-cooling system discussed and optically characterized above, we demonstrate integrated-photonics-based polarization-gradient cooling of a trapped ion for the first time.

For this demonstration, we utilize a planar electronic-photonic trap chip containing a trap site with one of the single-fiber-input TE-TE systems. This system contains two of the nominal TE gratings, designed to form beams with an intersection point 50 μm over the surface of the trap chip. We install this trap chip in a cryogenic vacuum system [42,43] for the purpose of achieving ultrahigh vacuum through cryopumping. The cryogenic environment is equipped with a fiber-feedthrough system to allow for coupling of light onto the chip from an off-chip 422-nm-wavelength laser located outside of the cryogenic system.

Once the electronic-photonic trap chip is installed, we load $^{88}$Sr$^+$ ions by photoionizing neutral Sr atoms from a remote precooled source [42,43]. Upon loading an ion, we shuttle the ion to a polarization-gradient-cooling zone located approximately 240 μm away along the axial dimension of the trap chip. This zone contains the two orthogonally-oriented TE gratings comprising our single-fiber-input TE-TE polarization-gradient-cooling system, as depicted in Fig. 8a.

As characterized above, the fabricated gratings in our polarization-gradient-cooling system exhibit an offset from their designed height of intersection, causing the highest-intensity regions of the emitted beams to intersect approximately 9 μm higher than the target height of 50 μm. As this intersection is far from the optimal operating range around the null of the trap chip's integrated radio-frequency Paul trap, ions positioned at the dual-beam intersection are not well confined. Therefore, we confine ions within the trap null at a height of 50 μm and implement polarization-gradient cooling at the intersection of lower-intensity portions of the beams, as depicted in Fig. 8b.

Once the ion reaches its target position, we power the gratings in the polarization-gradient-cooling system and observe a unique feature of the resulting polarization gradient: the interferometric stability offered by the single-fiber-input system with splitting performed on chip allows for the formation of a phase-stable polarization gradient, akin to previously demonstrated integrated-photonics-based phase-stable standing waves [44]. In this section, we present a brief summary of the parameter optimization and cooling performance attained using the inherent phase stability of this system. We present a more thorough contextualization, discussion, and demonstration of ion dynamics within this phase-stable gradient in an accompanying publication [31].

The passive phase stability of our system facilitates several functionalities, including spatial mapping of the polarization gradient, as depicted in Fig. 8b. While similar measurements have previously been demonstrated, they have relied on techniques such as stroboscopic sampling to map a running-wave gradient [45]. To map the polarization gradient of our system, we use a measurement scheme based on selective state preparation, described in [31]. In summary, we apply a polarization-gradient pulse, resulting in the spatially-dependent preparation of the ion in one of the ground-state Zeeman sublevels. Then, we selectively excite the ion along the $S_{1/2} \rightarrow D_{5/2}$ transition and use the ion's fluorescence to evaluate the sublevel occupation (a process often referred to as electron shelving [46]). Based on the population in each Zeeman sublevel, we determine the local polarization. Despite the fabrication-induced offset in the height of intersection of our beams, we find that our polarization gradient closely matches the theoretically-anticipated gradient calculated in [31].

Beyond facilitating polarization-gradient mapping, the phase stability of our system allows us to choose a specific phase of the polarization gradient and thoroughly study the cooling/heating dynamics over a range of system parameters. In Fig. 8c, we plot the time dynamics of the cooling process in this phase-stable polarization gradient for one parameter set. Specifically, we plot the average motional state ($\langle n \rangle$) of the 1-MHz axial mode versus time for a trapped $^{88}$Sr$^+$ ion initialized with Doppler cooling. The exponential decay of the motion with time constant $50.627 \pm 0.003$ μs and steady state $\langle n \rangle$ of $2.31 \pm 0.32$ demonstrates successful polarization-gradient cooling to significantly below the Doppler limit ($\langle n \rangle \approx 10$). In [31], we demonstrate further results corresponding to parameter sweeps over ion position, frequency detuning, and optical power.

Finally, in this work, we characterize the stability of our ion/polarization-gradient system by measuring the relative position between the ion and polarization gradient over the course of approximately 15 hours. At each point in time, we fit a sine function to both the $g_{-1/2}$ and $g_{1/2}$ population data collected in state-preparation experiments while scanning the ion position along the trap axis through the polarization gradient. We then convert the relative phase of the sine functions to a positional translation of the ion relative to the polarization gradient. Using this process, we measure a maximum relative drift of approximately 10 nm/hour immediately following charging of the electronic-photonic trap-chip surface induced during ion loading. As the trap surface discharges, the displacement rate of the polarization gradient exponentially decays, as depicted in Fig. 8d. To compensate for this displacement, we use a closed-loop feedback scheme detailed in [31] to adjust the position of the ion with respect to the polarization gradient. This stabilizes the relative position between the ion and the polarization gradient to < 10 nm (i.e. approximately $0.02\lambda_{eff}$, where $\lambda_{eff}$ is the periodicity of the polarization gradient as computed in [31]).

## Discussion

In this paper, we designed and experimentally demonstrated key polarization-diverse integrated-photonics devices and utilized them to implement a variety of integrated-photonics-based polarization-gradient-cooling systems, culminating in the first experimental demonstration of polarization-gradient cooling of a trapped ion by an integrated-photonics-based system.

First, we discussed the mechanisms underlying polarization-gradient cooling and introduced several integrated-photonics-based architectures capable of performing polarization-gradient cooling. Second, we designed and experimentally demonstrated the key integrated-photonics routing and grating-emitter devices required to implement these architectures. Third, we incorporated these devices into, and successfully experimentally demonstrated, multiple integrated-photonics-based polarization-gradient-cooling systems. Finally, we implemented one of these systems within an integrated-photonics-based ion-trap chip and used it to demonstrate the first polarization-gradient cooling of a trapped ion using an integrated-photonics-based system.

By demonstrating polarization-gradient cooling using an integrated-photonics-based system, this work paves the way for higher-efficiency integrated-photonics-based trapped-ion computations. Moreover, in general, this work opens up the field of polarization-diverse integrated-photonics-based devices and systems for trapped ions, enabling future advanced operations for integrated-photonics-based trapped-ion platforms. Accordingly, this work has the potential to facilitate further advancements in such platforms that improve upon the functionality of free-space trapped-ion systems with the intrinsic scalability and stability of integrated photonics.

## Data Availability

Data is available upon reasonable request.

## Acknowledgements

The authors thank Daniel M. DeSantis for thoughtful discussions and the MIT Office of Research Computing and Data (ORCD) for providing high-performance-computing and consultation resources. This work is supported by a collaboration between the US DOE and other Agencies. This material is based upon work supported by the U.S. Department of Energy, Office of Science, National Quantum Information Science Research Centers, Quantum Systems Accelerator. Additional support is acknowledged from the NSF Quantum Leap Challenge Institute Hybrid Quantum Architectures and Networks (QLCI HQAN) (2016136), NSF Quantum Leap Challenge Institute Quantum Systems through Entangled Science and Engineering (QLCI Q-SEnSE) (2016244), MIT Center for Quantum Engineering (H98230-19-C-0292), NSF Graduate Research Fellowships Program (GRFP) (1122374), Department of Defense National Defense Science and Engineering Graduate (NDSEG) Fellowship, MIT Rolf G. Locher Endowed Fellowship, and MIT Frederick and Barbara Cronin Fellowship. This material is based upon work supported by the Department of Energy and the Under Secretary of Defense for Research and Engineering under Air Force Contract no. FA8702-15-D-0001. Any opinions, findings, conclusions, or recommendations expressed in this material are those of the author(s) and do not necessarily reflect the views of the Department of Energy or the Under Secretary of Defense for Research and Engineering.

## Conflict of Interest

The authors declare no competing financial interests.

## Contributions

S.M.C., A.H, M.N., R.S., and T.S. designed the integrated-photonics-based devices and systems. P.T.C. and T.M. laid out the devices and systems. D.K. led the chip fabrication. S.M.C., A.H., G.N.W., and M.N.

performed the photonics measurements. E.R.C., F.W.K., and S.M.C. performed the ion experiments. J.N., J.C., R.M., I.L.C., C.S.A., and C.D.B. supervised the project.

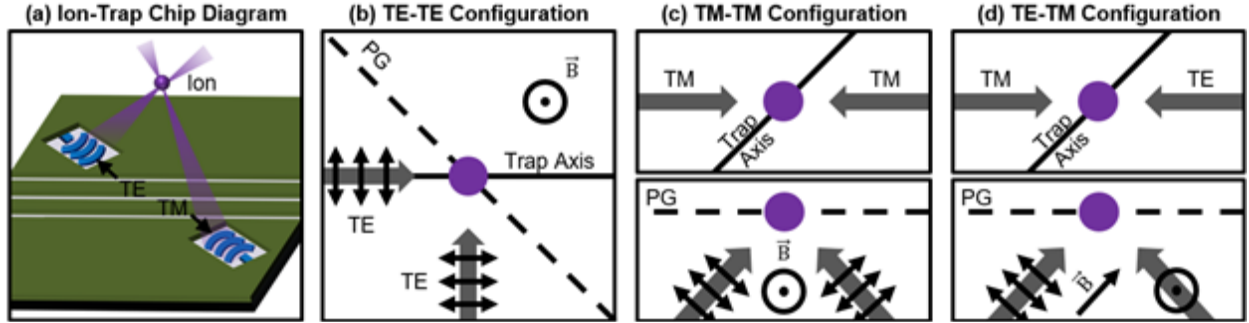

**Figure 1: Integrated-photonics-based polarization-gradient-cooling concept.** (a) Conceptual diagram of a planar electronic-photonic trap chip containing integrated-photonics gratings emitting optical beams upward towards a common ion location through surface-electrode windows (not to scale). (b-d) Conceptual diagrams of the polarization-gradient-cooling configurations explored in this work, depicting the beam, polarization-gradient (PG), and magnetic-field ($\vec{B}$) orientations. (b) Top-view conceptual diagram of a TE-TE polarization-gradient-cooling configuration. (c) Top-view (top) and side-view (bottom) conceptual diagrams of a TM-TM polarization-gradient-cooling configuration. (d) Top-view (top) and side-view (bottom) conceptual diagrams of a TE-TM polarization-gradient-cooling configuration.

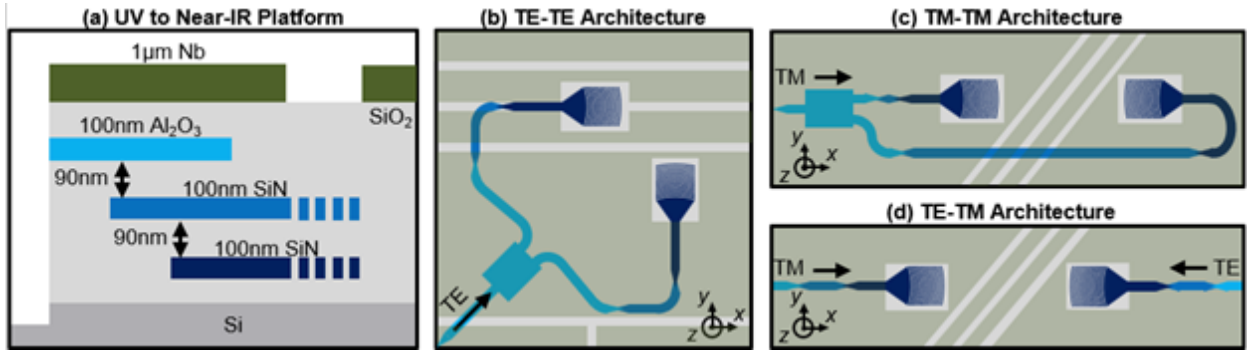

**Figure 2: Integrated-photonics-based polarization-gradient-cooling architecture implementation.** (a) Simplified schematic (not to scale) of the low-loss UV-to-near-IR 200-mm wafer-scale platform [15] used to fabricate the electronic-photonic trap chip utilized in this work, consisting of three photonic waveguiding layers in $Al_2O_3$ and SiN underneath an Nb surface-electrode layer. Top-view schematics (not to scale) of the (b) TE-TE, (c) TM-TM, and (d) TE-TM architectures designed and demonstrated in this work, illustrating the integrated-photonics devices implemented in $Al_2O_3$, single-layer SiN, and dual-layer SiN, as well as the alignment of the Nb surface electrodes with respect to the dual-layer-SiN grating emitters.

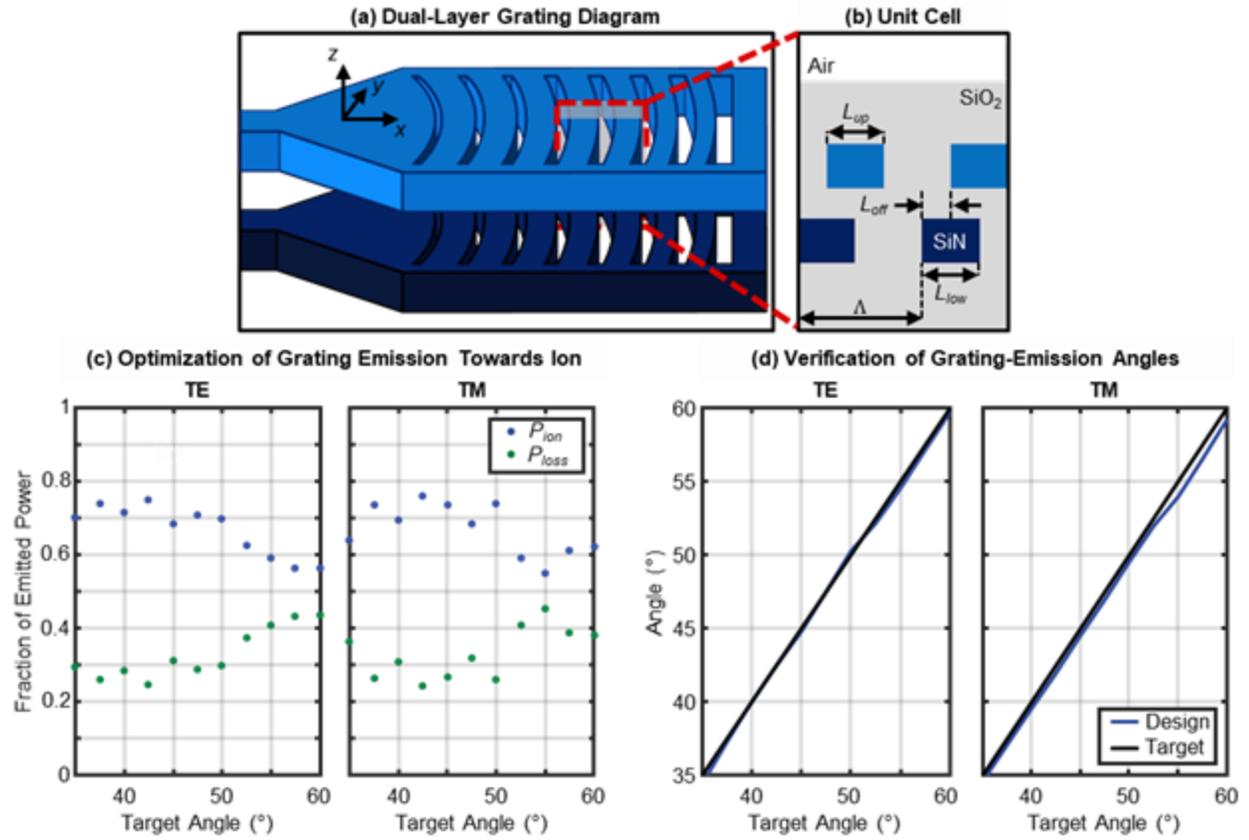

**Figure 3: Polarization-diverse integrated-photonics grating-emitter design process.** (a) Simplified diagram (not to scale) of our dual-layer-SiN integrated-photonics grating emitters, demonstrating apodization and curvature of the grating teeth for focusing-beam emission. (b) Cross-sectional view (not to scale) of two unit cells of the grating depicted in (a), highlighting our optimization parameters: local period ($\Lambda$), upper-tooth length ($L_{up}$), lower-tooth length ($L_{low}$), and offset between layers in the *x* dimension ($L_{off}$). TE- and TM-grating optimization-routine results, depicting (c) maximization of light emitted towards a desired ion location as a fraction of the total emitted light ($P_{ion}$) and minimization of light emitted in all other directions as a fraction of the total emitted light ($P_{loss}$) versus the target angle of emission, and (d) simulated optimized angles of emission along the grating versus the desired target angles.

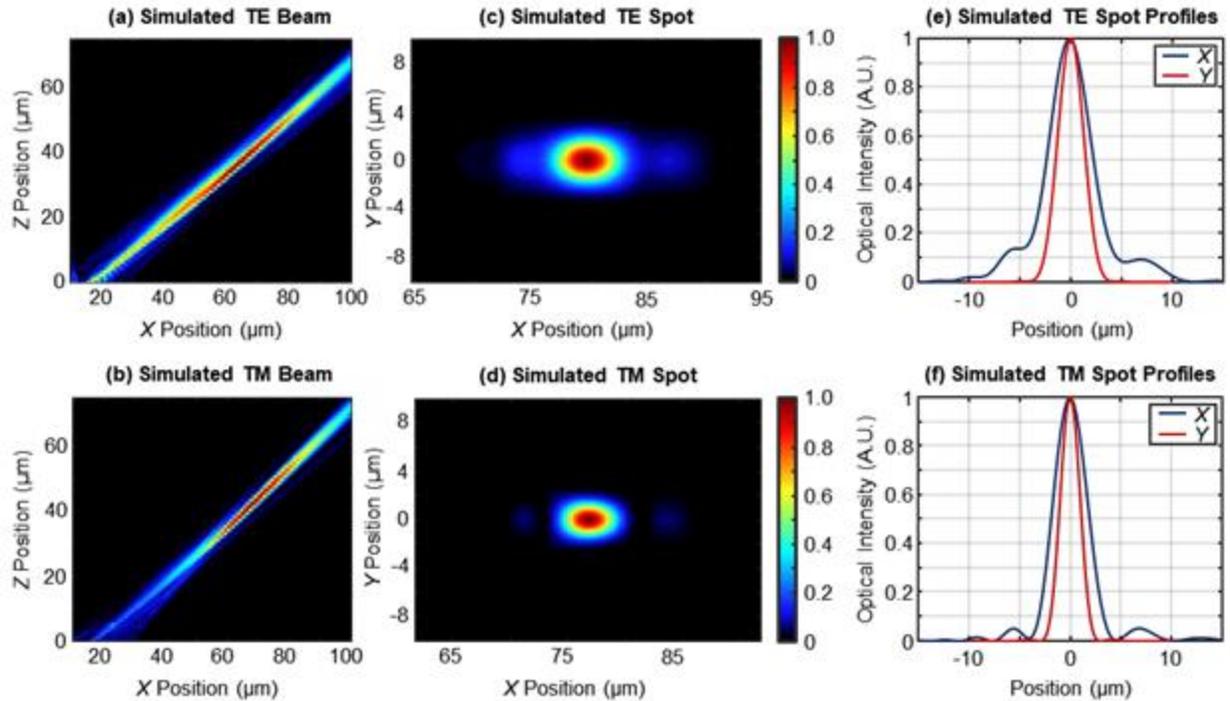

**Figure 4: Polarization-diverse integrated-photonics grating-emitter simulated performance.** Simulated *xz*-plane beam-intensity profiles for an example characteristic (a) TE grating (17 μm long and targeting $x$ = 79.7 μm) and (b) TM grating (17 μm long and targeting $x$ = 76.7 μm). Simulated spots in the *xy* plane at $z$ = 50 μm for the characteristic (c) TE and (d) TM grating. Simulated spot profiles in the *x* (blue) and *y* (red) dimension at $z$ = 50 μm for the characteristic (e) TE and (f) TM grating.

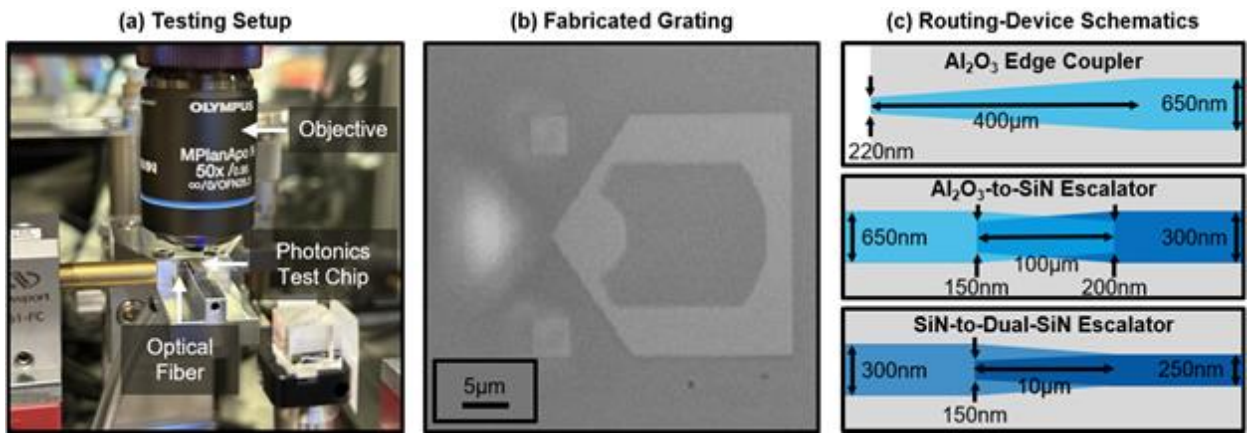

**Figure 5: Automated experimental setup and waveguide-layer routing devices.** (a) Photograph of the automated testing setup used to experimentally characterize the gratings and systems developed in this work, depicting an input optical fiber coupled to a photonics test chip mounted on a vacuum chuck underneath a 50X objective. (b) Micrograph of a fabricated grating imaged using the setup in (a). (c) Simplified top-view schematics (not to scale) of the integrated-photonics waveguide-based inverse-taper

fiber-to-chip edge coupler (top), Al$_2$O$_3$-to-SiN layer-transition escalator (center), and SiN-to-dual-SiN layer-transition escalator (bottom).

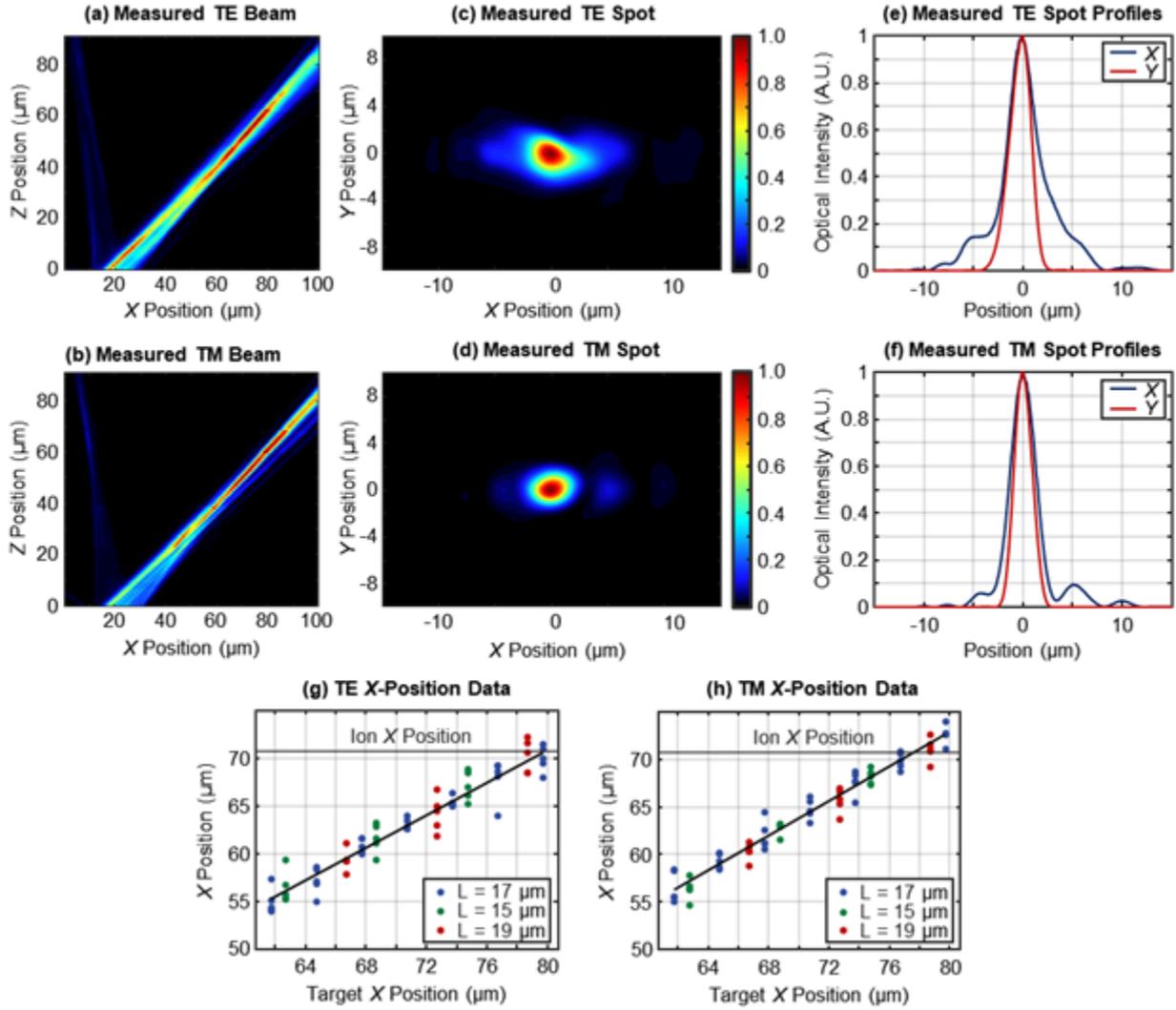

**Figure 6: Polarization-diverse integrated-photonics grating-emitter experimental results.** Experimentally-measured *xz*-plane beam-intensity profiles for a characteristic (a) TE grating (17 μm long and targeting $x = 79.7$ μm) and (b) TM grating (17 μm long and targeting $x = 76.7$ μm). Experimentally-measured spots in the *xy* plane at $z = 50$ μm for the characteristic (c) TE and (d) TM grating. Experimentally-measured spot profiles in the *x* (blue) and *y* (red) dimension at $z = 50$ μm for the characteristic (e) TE and (f) TM grating. Experimentally-measured versus target beam *x* positions at $z = 50$ μm for (g) 13 TE and (h) 13 TM grating variants from 5 chips with grating lengths of 17 μm (blue), 15 μm (green), and 19 μm (red) (total of 130 gratings measured).

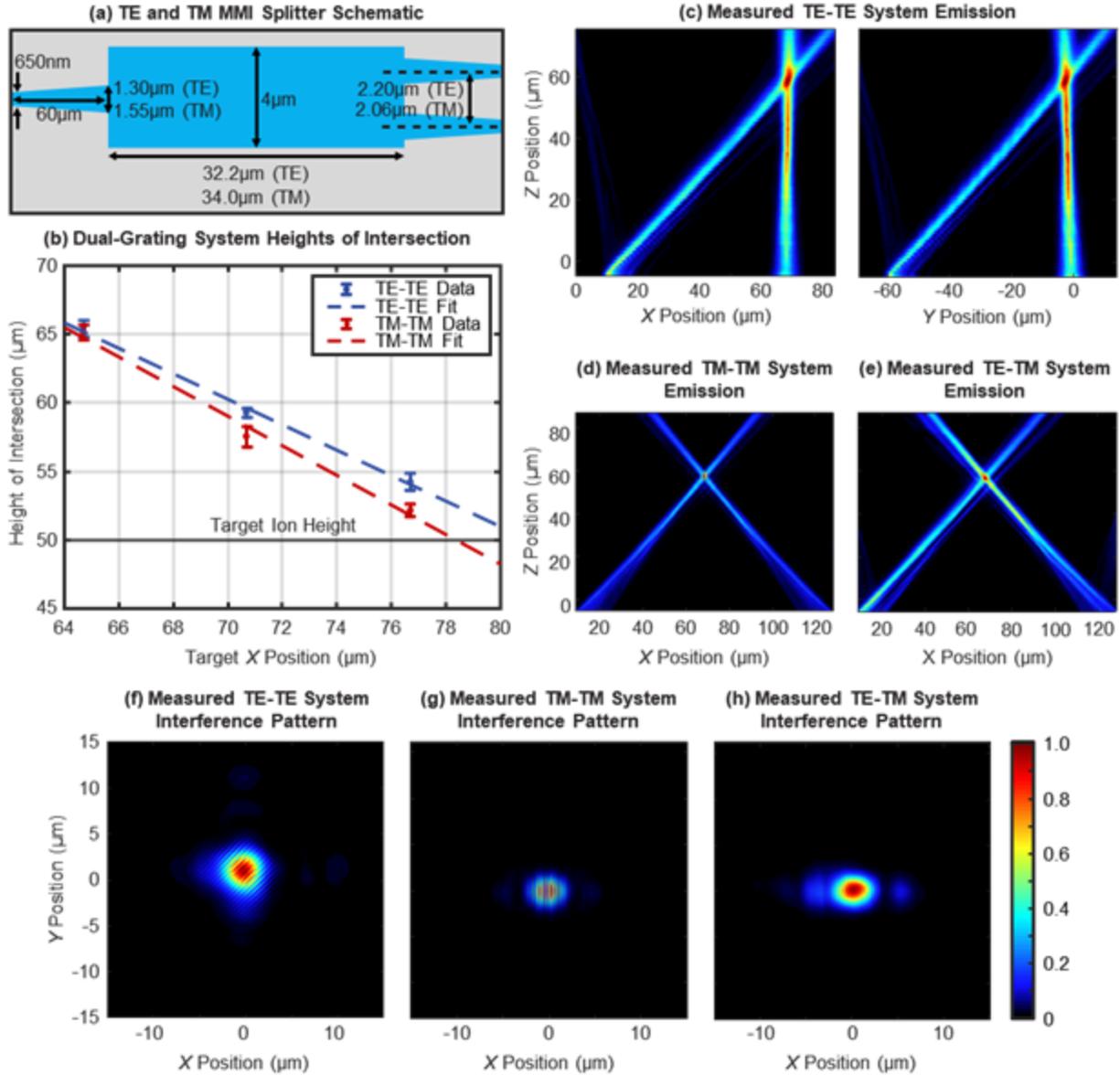

**Figure 7: Integrated-photonics-based polarization-gradient-cooling systems experimental results.** (a) Simplified schematic (not to scale) of the TE and the TM integrated-photonics $Al_2O_3$ multimode-interferometer (MMI) splitter with relevant dimensions labeled for each polarization. (b) Measured beam-intersection heights for the TE-TE (blue) and TM-TM (red) systems compared to the target ion height of $z$ = 50 μm (black) (total of 48 systems measured). (c) Experimentally-measured $xz$-plane (left) and $yz$-plane (right) emission profiles for a TE-TE system containing two nominal TE gratings. Experimentally-measured $xz$-plane emission profiles for (d) a TM-TM and (e) a TE-TM system. Experimentally-measured $xy$-plane views of the emission from the (f) TE-TE system at $z$ = 60 μm, (g) TM-TM system at $z$ = 59 μm, and (h) TE-TM system at $z$ = 58 μm, showing the resulting dual-beam interference patterns.

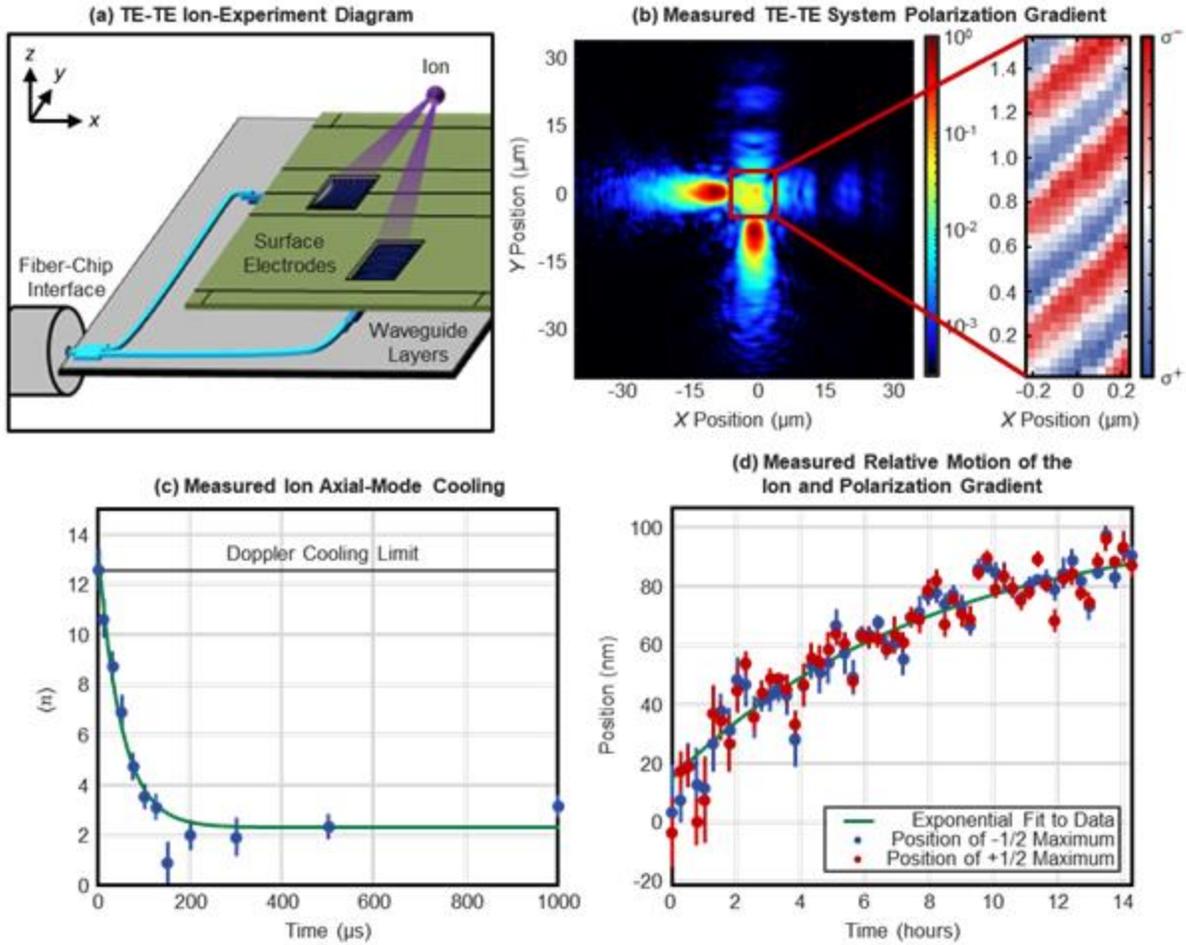

**Figure 8: Integrated-photonics-based polarization-gradient-cooling ion results.** (a) Conceptual diagram of integrated-photonics-based polarization-gradient cooling of a trapped ion, depicting light coupled onto a chip from an optical fiber, routed through integrated-photonics waveguiding layers, and emitted upward towards a common ion location through surface-electrode windows (not to scale). (b) Experimentally-measured *xy*-plane views at $z = 50$ μm of the emission from the nominal TE-TE system (in log scale), with an inset depicting the resulting polarization gradient, characterized using fluorescence from a trapped $^{88}$Sr$^+$ ion. (c) Experimentally-measured average motional state ($\langle n \rangle$) of the 1-MHz axial mode versus time for a $^{88}$Sr$^+$ ion in the generated polarization gradient, demonstrating successful polarization-gradient cooling to significantly below the Doppler cooling limit. (d) Experimentally-measured position versus time of the $g_{1/2}$ population maximum (red) and $g_{-1/2}$ population maximum (blue) for a $^{88}$Sr$^+$ ion in the generated polarization gradient, demonstrating maximal motion of approximately 10 nm/hour and confirming the passive phase stability of the integrated-photonics-based polarization gradient.

23